# Mechanical Fatigue on Gold MEMS Devices: Experimental Results

G. De Pasquale, A. Soma', A. Ballestra
Department of Mechanics, Polytechnic of Torino
Corso Duca degli Abruzzi 24, 10129 Torino, Italy.

*Abstract* – **The effect of mechanical fatigue on structural performances of gold devices is investigated. The pull-in voltage of special testing micro-systems is monitored during the cyclical load application. The mechanical collapse is identified as a dramatic loss of mechanical strength of the specimen. The fatigue limit is estimated through the** *stair-case* **method by means of the pull-in voltage detections. Measurements are performed by means of the optical interferometric technique.**

## I. INTRODUCTION

Reliability of MEMS increased its relevancy together with the spread of micro-devices applications in the every day life. The functioning of micro-fluidic bioMEMS, for example, can have a significant impact on the vitality of patients; aerospace applications impose more stringent conditions on the performance and reliability to limit repairs and replacements, etc.

About structural reliability, it is necessary to examine effects of process parameters, geometry, loads and working environmental conditions, which are all aspects whose knowledge is at present very scarce. MEMS reliability deals with materials but also with the reliability of their mechanical components. Moreover MEMS materials are generally used as thin films and this makes mechanical characterisation more complicated and at present no standard procedures are available for the extraction of fatigue properties and for the detection of all factors influencing fatigue behaviour as environmental humidity [1], building processes, etc. Also the utilization of multiple test structures for testing can introduce a device-dependency of results [2]. Utilization of materials in thin films generally enhances material properties of the material itself with respect to traditional applications, but still the determination of mechanical properties at microscopic scale is remarkably uncertain. This is due to the variation of material characteristics among different producers, among wafers of the same producer and even among structures on the same wafer. At present, most part of the researches about material reliability deal with polysilicon [3], since it is the most used material both for mechanical and electric parts of MEMS; it can be considered as a material with a dominant brittle behaviour. About test methodologies, some different examples are reported [4 - 9].

Present work is based on gold devices representing an example of ductile material to which essentially corresponds a lack of information in the literature. The mechanical strength of the structure is affected by the damage producing inside the material during the oscillation; the vibration test produces a cyclical load on the structure and cyclical concentrated stresses which are at the basis of the fatigue phenomenon. The consequence of this is the progressive lost of performances till the final failure that can be enhanced by the pull-in voltage decrease.

## II. SPECIMENS AND TESTING PROCEDURE

The shape of test structures used is represented in figure 1 and the relative geometrical nominal dimensions are reported in table I; the device contains both the actuation part (represented by the perforated plate) and the specimen (the beam); this peculiarity is quite common in micro-systems for fatigue tests representing an example of the *embedding* inside a complete micro fatigue-machine. Described devices are designed *ad-hoc* by means of FE models in order to generate a shear stress inside the specimen [10, 11]. It is possible to observe that the beam deflection is much smaller than the plate length; this produces a rotation of the un-clamped extremity of the specimen approximately equal to 0.4 degrees, allowing to assume a vertical movement of it without rotation, corresponding to a pure shear excitation.

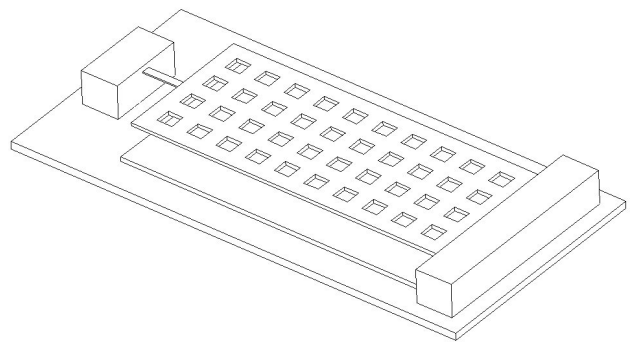

Fig. 1. Geometrical shape of the micro fatigue-machine (the gap thickness under the plate is enlarged).

The ITC-IRST [12] building technology is used; the lower electrode is realized by polysilicon, suspended parts are obtained through a sacrificial layer upon which a single (for the specimen) or a double (for the plate) gold layer is deposited. Nominal material parameters are $E = 98.5\,\text{GPa}$, $\nu = 0.42$, $\rho = 19.32 \cdot 10^{-15}\,\text{kg}/\mu\text{m}^3$. The value of static yield stress is studied in [13] for gold material specimens and different thickness and width. In [13] it is shown the strong influence of the dimension of the specimen on the tests results. This effect can be adressed to the intensification factor of the mechanical stress and to local effects. In the present work the failure of the specimen has been monitored by using global parameters such as pull-in voltage instead of local parameters as the mechanical stress.

Further work on identification of mechanical stress distribution on FE models will give the possibility to compare the stress local effect with the pull-in voltage variation.

The pull-in voltage results to be the most significant parameter able to monitor the cumulative damage of the material for the testing devices used. The pull-in of the not damaged structure is measured by supplying a DC voltage which is increased step by step; the cyclical load is generated by means of an AC voltage at the frequency of 20kHz. The excitation at resonance (about 28kHz measured) is avoided because of two main reasons: 1) the resonant amplification determines additional difficulties evaluating internal stresses in the material, 2) the progressive damage of the material could shift the resonant peak or modify the quality factor considering eventual changes of the structural stiffness and damping. Repetitive successive pull-in detections are performed and stored during the fatigue loading at fixed time intervals. The pull-in positioning is verified directly by an optical control through the interferometric microscope exploiting the translation of the fringes. The ZoomSurf3D (Fogale Nanotech) microscope [14] is used.

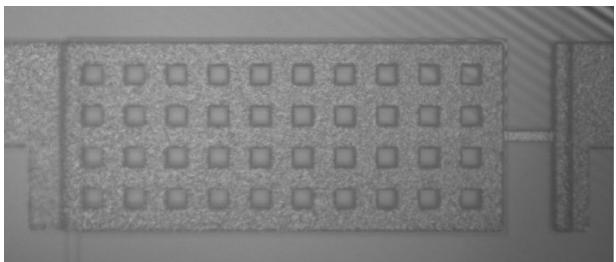

Fig. 2. Microscope image of the micro fatigue-machine.

TABLE I
NOMINAL DIMENSIONS (IN MICRONS) OF THE MICRO FATIGUE-MACHINE

| specimen length | 50 | holes side | 20 |
|---|---|---|---|
| specimen width | 10 | holes interspace | 20 |
| specimen thickness | 1.8 | number of holes | 10x4 |
| plate length | 420 | lower electrode length | 420 |
| plate width | 180 | lower electrode width | 460 |
| plate thickness | 4.8 | gap thickness | 3 |

## II. CYCLICAL EXCITATION VOLTAGE

The main purpose of a fatigue analysis is that the results produced by tests on specimens can be used for more general cases; this requires to extract from the specific test some fundamental quantities, as the maximum stress ($\sigma_{max}$), the minimum stress ($\sigma_{min}$), the mean stress ($\sigma_m = (\sigma_{max}+\sigma_{min})/2$) and the alternate stress ($\sigma_a = (\sigma_{max}-\sigma_{min})/2$) of the loading cycle, the stress ratio ($R = \sigma_{min}/\sigma_{max}$), the fatigue limit ($\sigma_D$), and the number of cycles to failure (N).

The number of cycles of excitation needs to be carefully considered in the conversion from the frequency of the supply voltage to the loading cycles. The specimen oscillation is generated by the potential difference between the plate and the lower electrode that produces an attractive force; as reported in the diagram of figure 3, two periods $T_L$ of loading (or two cycles of actuation) correspond to one period $T_V$ of the alternate voltage. This is due to the present mode of deflection that not allows an oscillation around the undeformed position of the structure; the described situation takes place for all suspended structures in MEMS actuated by an alternate voltage. The resulting number of cycles $N_L$ of loading (and cycles of deflection) is related to the number of cycles of the alternate voltage $N_V$ by the relation

$$N_L = 2N_V. \qquad (1)$$

It is evident that some portions of the specimen are always subjected to a local compression loading (here R is constant and equal to $\infty$) while symmetrical portions are always subjected to a local traction loading (here R is constant and equal to 0). These last regions are expected to be responsible of cracks nucleation. After these considerations all the fatigue loading parameters can be easily calculated from the value of the alternate load.

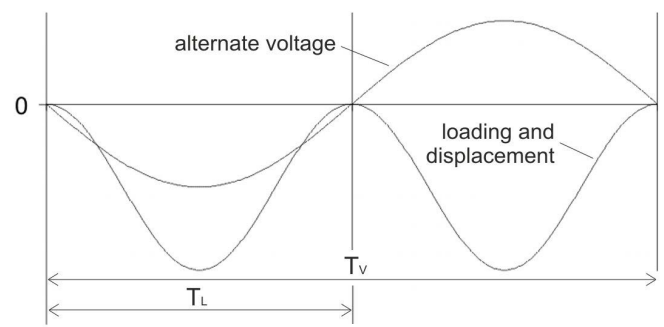

Fig. 3. Qualitative actuation voltage cycle and fatigue load cycle.

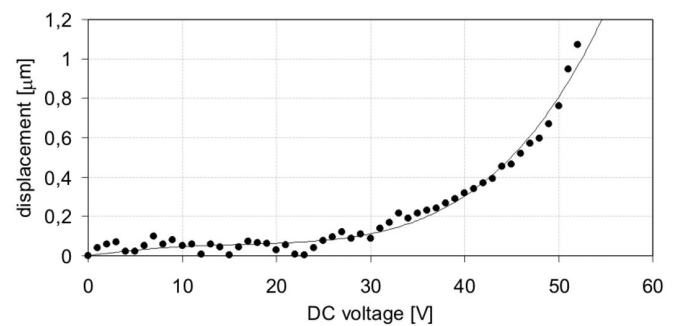

Fig. 4. Conversion curve from alternate voltage to alternate equivalent stress.

The static displacement of the specimen is measured by the interferometric microscope at different levels of DC voltages and the curve reported in figure 4 is obtained.

This static acquisition is made assuming the absence of amplification of the vibration in the dynamic condition; this is reasonable considering that the excitation frequency is quite distant to the resonance.

## III. Experimental Results

The fatigue limit is estimated by the *stair case* method [15]; the reference number of cycles is fixed to $2 \cdot 10^6$, the load step is $1V_a$ and four load levels ranging from 12 to $15V_a$ are selected. The results of stair case tests on six specimens are reported in table II, where the failure is indicated as 1 and the non-failure as 0. The fatigue limit at 50% of failure probability results to be $V_D = 13V$; similarly the fatigue limits at 10 and 90% of failure probability are calculated as $V_{D-10\%} = 12.3V$ and $V_{D-90\%} = 13.7V$.

TABLE II
STAIR-CASE RESULTS FOR $V_D$ ESTIMATION

| loading level | specimen | | | | | |
|---|---|---|---|---|---|---|
| | 1 | 2 | 3 | 4 | 5 | 6 |
| 15 V | 1 | | | | | |
| 14 V | | 1 | | 1 | | |
| 13 V | | | 0 | | 1 | |
| 12 V | | | | | | 0 |

Next figure 6 reports the measured values of pull-in voltages for different specimens during the accumulation of loading cycles. Only collapsed specimens are represented here; the failure is assumed to happen at the moment when the pull-in voltage registers a strong reduction of its amount. The figure 7 instead represents specimens that collapsed instantaneously because of the high level of fatigue loading (21 or 22.5V) and specimens that do not collapse as excited at a loading (10, 12 or 13V) equal or lower than the fatigue limit.

The Wöhler diagram resulting from reported curves is represented in figure 5 where the limit value is reported as pull-in voltage.

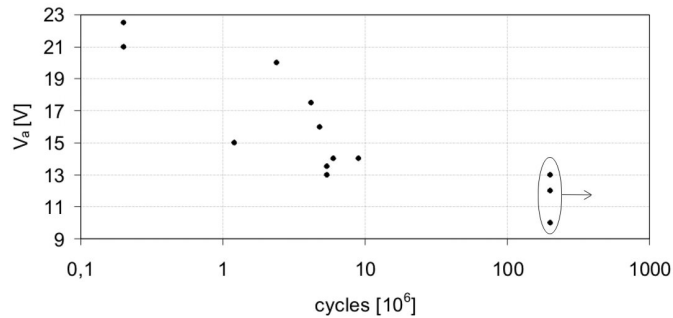

Fig. 5. Wöhler diagram of fatigue tests curves reported.

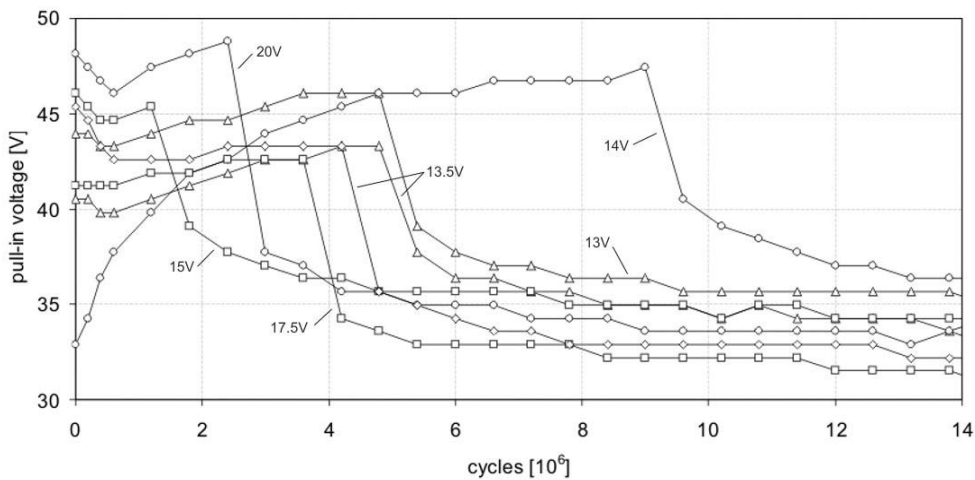

Fig. 6. Experimental variation of the pull-in voltage with respect to the loading cycles; the fatigue loading is provided by the excitation voltage at 20kHz at the amplitude indicated for each curve.

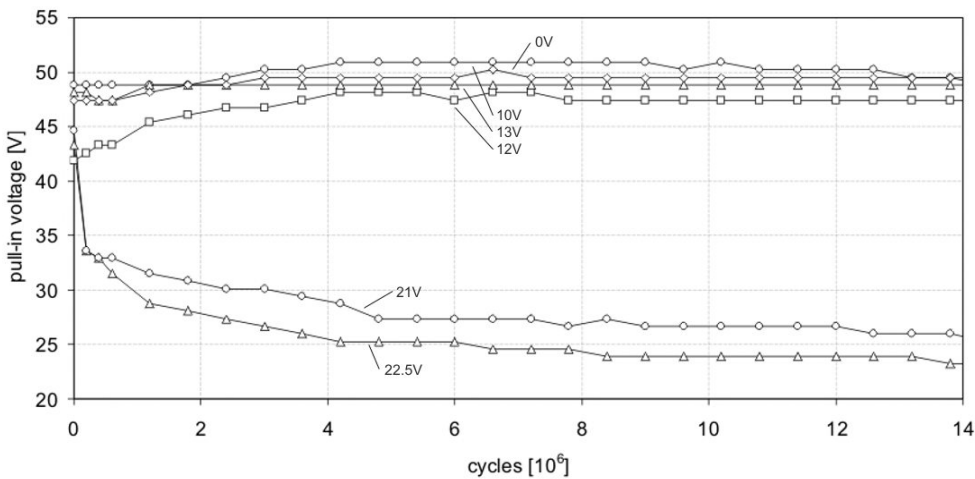

Fig. 7. Experimental variation of the pull-in voltage with respect to the loading cycles; specimens collapsed immediately or non-collapsed are reported.

## IV. DISCUSSIONS

Presented fatigue tests use the pull-in voltage as the parameter to monitor the progressive material damage; resonance frequency and quality factor have been used in the literature at this purpose, but they are observed not to be effective on devices used here because their variation is not appreciable. The pull-in detection determines a stress level in the material quite high if compared with the stress generated during the fatigue loading; by the FE model is estimated that the material reaches the yield condition locally. However this one-cycle actuation is not responsible for the plastic failure, as demonstrated by the fatigue loading at 0V reported in figure 7; here the specimen is only subjected to a series of successive pull-in actuations which actually not affect its strength.

All reported curves refers to stress-imposed fatigue tests; it is observed that, for high values of $V_a$ (and $\sigma_a$), the pull-in voltage curve can decrease during the fatigue loading till to reach the value of the alternate voltage of actuation. This condition causes a completely different type of vibration and the specimen results to be excited to the pull-in at each cycle; so the fatigue test becomes displacement-imposed and no more results comparable to the others can be obtained.

The mechanical fatigue collapse can by divided into three separate phases for simplicity: the crack *nucleation*, the crack *stable propagation* and the final *failure*. The last phase is not intended here as the complete rupture of the specimen, but as the dramatic loss of its mechanical strength. The first and the second phases are expected in the portions of the material which are subjected to an alternate traction loading (where R = 0).

The motion of dislocations in ductile materials takes place at cyclical stress levels much smaller than yield stress, resulting in a sliding movement of crystal planes near the surface of the loaded component. This process, influenced by the grain size, the lattice geometry, the temperature, etc. determines the superficial nucleation of the crack and its successive propagation inside the material. The additional effect of dislocations is to increase the local material strength (*work hardening*) when they accumulates at the grain borders or near a plasticized zone of the material; this is the situation resulting from the experimental curves reported in figure 6, where a sensible strength increase is evident before the collapse. The regions of the material that reach the plasticity condition after the pull-in detection determine a sort of barrier at the motion of dislocations in that zone.

The temperature influence on fatigue collapse is also observed during the tests; this parameter in fact strongly influences the motion of dislocations. In particular, when the time of the cyclical loading is increased between two successive pull-in detections, the pull-in value results to be quite smaller than the expected one; this is due to the additional mechanical heat transfer towards the material during the oscillation. The frequency of excitation also influences the temperature of the specimen.

Other factors affecting the described fatigue behaviour are the geometrical discontinuities (like the right angles used to connect the specimen with the plate and the support) which acts as stress multipliers, the typology of the applied stress and the building processes (especially related to thermal treatments and superficial roughness).

## V. CONCLUSIONS

Mechanical fatigue behaviour of gold material applied to MEMS structures is studied by a special micro fatigue-machine; the pull-in voltage is used to monitor the damage progress and the resulting Wöhler diagram is traced in terms of alternate voltage. The fatigue limit is estimated through the stair-case method.

A further work will produce the developement of FE models to obtain a quantitative conversion from the fatigue parameters expressed in volts to material stress. The superficial analysis of damaged specimens by a SEM will recognize characteristic degradation features. The number of tests needs to be increased in order to give a statistically treatment to the results and other strategies to investigate the material damage could be adopted, as the detection of the specimen electrical resistance variation. Additional improvements to the analysis can be obtained by studying different specimen shapes or by adopting different actuation strategies.